\documentclass[runningheads]{llncs}
\usepackage[T1]{fontenc}
\usepackage{cite}
\usepackage{algorithmic}
\usepackage{graphicx}
\usepackage{textcomp}
\usepackage{xcolor}

\newcommand{\figref}[1]{Fig.~\ref{#1}}
\usepackage{hyperref}
\usepackage{url}
\usepackage{siunitx}
\usepackage{xspace}

\usepackage[normalem]{ulem}

\DeclareSIUnit{\flop}{flop}
\DeclareSIUnit{\gCOeq}{g\,CO_{2}eq}
\DeclareSIUnit{\tCOeq}{t\,CO_{2}eq}
\DeclareSIUnit{\watth}{Wh}
\DeclareSIUnit{\gpuhour}{hGPU}

\begin{document}

\newcommand{\kuma}{Kuma\xspace}

\title{Green500: Adventure in Qualifying a Cluster}
%
%
\author{Nicolas Richart\inst{1}\orcidID{0000-0002-1463-4405} \and
Julia Paolini\inst{2}\orcidID{0009-0004-1283-9943} \and
Xavier Ouvrard\inst{2}\orcidID{0000-0003-4054-2771} \and
Manuel Cubero-Castan\inst{3}\orcidID{0009-0005-3152-6915} \and
Gilles Fourestey\inst{1}\orcidID{0000-0001-7217-2977}}
\authorrunning{N. Richart et al.}
%
\institute{SCITAS, École Polytechnique Fédérale de Lausanne (EPFL),
1015 Lausanne, Switzerland \email{nicolas.richart@epfl.ch} \and
EcoCloud, École Polytechnique Fédérale de Lausanne (EPFL), 1015 Lausanne, Switzerland \and
Sustainability, École Polytechnique Fédérale de Lausanne (EPFL), 1015 Lausanne, Switzerland
}
%
\maketitle              

\begin{abstract}
  Energy efficiency has become a key metric for evaluating High Performance
  Computing (HPC) systems, as reflected by the Green500 ranking. However, energy
  efficiency alone does not capture the full environmental impact of an HPC
  cluster. Using the Kuma cluster at EPFL, we report on two Green500 measurement
  campaigns at Level~2 and Level~3, and extend this analysis to a full carbon
  footprint assessment using the MITSI life-cycle based methodology. GPU power
  capping experiments show that energy efficiency peaks at \qty{450}{\watt},
  achieving an 11\% performance loss offset by a 20\% reduction in energy
  consumption. When embodied emissions are accounted for, the optimal power cap
  for carbon efficiency shifts to \qty{500}{\watt} under Switzerland's
  low-carbon electrical grid, revealing a tension between minimizing usage and
  embodied contributions. We discuss the challenges of generalizing this
  approach into a broader carbon-based HPC ranking, and argue that moving beyond
  energy efficiency is both necessary and feasible.

\end{abstract}

\section{State of the Art}

High Performance Computing (HPC) systems have evolved through successive
architectural paradigms, from vector supercomputers to massively parallel
processing systems, commodity clusters, and heterogeneous accelerator-based
nodes. Comparing these infrastructures became key, and the introduction of the
LINPACK benchmark \cite{dongarra1979linpack} enabled the TOP500
list\cite{top500-1993}, first published in 1993, to provide a standardized
mechanism to track long-term progress in HPC. TOP500, which ranks the first 500
HPC systems, privileges measured rather than peak performance, highlighting the
importance of scalable infrastructures, efficient interconnects and optimized
numerical libraries.

While HPL---the implementation of LINPACK---allows fine tuning of parameters
across a wide range of system sizes\cite{strohmaier2015top500}, it focuses
solely on floating-point performance and cannot evaluate other aspects such as
bandwidth or memory access. Additionally, it focuses on performance-driven
clusters rather than sustainability-driven ones.

As early as 2007, several actors were starting to push the idea that another
ranking was needed in addition to the TOP500. In \cite{feng-2007}, the authors were
already mentioning how performance-at-any-power-cost in HPC was not sustainable,
and put the foundations of the Green500. In the Green500\cite{green500-2013},
the HPC clusters that the TOP500 ranks solely on performance are instead ranked
by energy efficiency, measured in \unit{{\giga\flop\per\second}\per\watt}. The
first Green500 ranking was published in 2013. Since then the efficiency of HPC
systems has doubled every two years\cite{adhinarayanan2025looking}.

While alternative benchmarks such as HPL-MxP\cite{dongarra2025hpl,HPL-mixed},
HPL-AI\cite{kudo2020implementation}, HPCG\cite{heroux2013hpcg}, and
STREAM\cite{mccalpin1995stream} have emerged to address HPL's limitations, we
focus solely here on Green500, using HPL on the \kuma cluster\cite{SCITAS-Kuma}.

Focusing on energy efficiency should be seen as a first step toward
sustainability. Assessing the full environmental impact of HPC clusters requires
a more holistic approach \cite{kline2016holistically}, and in particular
evaluating the carbon footprint of such installations \cite{li2023toward}. In
particular, the whole life-cycle of the equipment has to be taken into account,
including the embodied emissions (those arising from manufacturing,
transportation, and disposal)\cite{gupta2021chasing}. We base our evaluation on
a new methodology that has been developed at EPFL, a collaboration between the
Sustainability department, the Scientific IT and Application Support (SCITAS)
platform and the EcoCloud research center.

\section{Green500}\label{Green500}

\subsection{Experimental Set-up}\label{sec:setup}

The Green500 power measurement methodology\cite{green500-methodology} defines
three quality levels across different aspects, granularity, machine fraction,
included subsystems, and instrumentation sampling rate, accuracy and location.

\subsubsection{The measured system}

The experiments described in this paper were conducted on the \kuma
cluster\cite{SCITAS-Kuma}. \kuma is composed of 84 compute nodes, each equipped
with 4 NVIDIA H100 SXM5 GPUs (\qty{94}{\giga\byte} HBM2e per GPU, interconnected
via NVLink at \qty{900}{\giga\byte\per\second}), 64 CPU cores, and
\qty{384}{\giga\byte} of DDR5 RAM. Each node is connected to the network fabric
through two \qty{200}{\giga\bit\per\second} InfiniBand HDR links. The nodes are
distributed across 10~racks; 9~racks of compute nodes and 1~administration rack
with the front and administration nodes and the shared storage. There is also an
extension with 20 similar nodes equipped with 8 NVIDIA L40s GPUs each instead of
the H100.

The GPUs are direct-to-chip liquid cooled while the CPUs are air cooled, with
the GPU heat transferred to the node's air cooling loop; active rear-door heat
exchangers use fans to help extract warm air from the racks. The recovered heat
is fed into EPFL's central heating system (using on heat pumps) during winter;
during summer water pumped from Lake Geneva is used to cool the rear-door
cooling loop.

For the Green500 submission, 80 out of the 84 nodes were used, as 4 nodes were
unavailable at the time of the measurements; the L40s nodes were turned off for
all the measurements. HPL was configured with a problem size
$N = 1{,}943{,}040$, a block size $NB = 768$, and a process grid of
$16 \times 20$ ($P \times Q$) for the 80 nodes runs. Single-node validation runs
used $N = 217{,}088$, $NB = 1{,}024$, and a $2 \times 2$ process grid.

\subsubsection{Level 2 -- List of November 2024}

For the Level~2 submission, power was measured using the Power Distribution
Units (PDUs) installed in each rack. The PDU power readings were collected via
SNMP and exported to a Prometheus time-series database using
\texttt{snmp\_exporter}, with Grafana used for visualization and data export.
The PDU readings provide per-rack active power at 1-second intervals, based on
sampling at \qty{4}{\kilo\hertz}, covering all 10 racks including the compute
nodes, the network, the administration and front nodes, and the shared storage.

Three GPU power capping configurations were tested: \qty{350}{\watt},
\qty{500}{\watt}, and \qty{700}{\watt}. For each configuration, HPL was run on
the full 80-node partition and power was recorded during both the core
computation phase (HPL time) and the full run (full SLURM job time). The nodes
used NVIDIA driver version 535.154.05.

\begin{table}[ht]
  \centering
  \begin{tabular}{lcccc}
    \hline
    Power cap & Perf. & Core Energy & Core Avg. Power & Energy Efficiency \\
              & (\unit{\peta\flop\per\second}) & (\unit{\kilo\watth}) & (\unit{\kilo\watt}) & (\unit{{\giga\flop\per\second}\per\watt}) \\
    \hline
    \qty{700}{\watt} & 11.95 & 31.2 & 274.4 & 43.555 \\
    \qty{500}{\watt} & 11.66 & 24.8 & 212.2 & 54.959 \\
    \qty{350}{\watt} &  8.85 & 35.6 & 167.0 & 52.995 \\
    \hline
  \end{tabular}
  \newline
  \caption{Level~2 results (PDU-based measurements)}\label{tab:level2}
\end{table}

\subsubsection{Level 3 -- List of November 2025}

The Level~3 requires energy-based measurements of the entire system at a sample
rate of \qty{5}{\kilo\hertz}. For this, power meters were installed on the
electrical distribution rails on the top of the racks. The compute nodes are
spread over 4 power rails (Rail~1 to Rail~4), each line having 3 phases, two of
which are connected through Uninterruptible Power Supplies (UPS). Different types
of power meters were used due to local availability: two A.~Eberle PQ-Box~50 on
Rail~1 and Rail~2, one A.~Eberle PQ-Box~150 on Rail~3, and one Fluke~1738 on
Rail~4. These instruments deliver
averaged measures at \qty{1}{\second} intervals and capture both instantaneous
power and cumulative energy. The PDU-based monitoring from Level~2 was
maintained in parallel, allowing a direct comparison between the two measurement
approaches.

Before the measurement campaign for the Level~3 submission, the cluster was
updated to RedHat 9.4 and the NVIDIA driver to version 560.35.03-1.
\subsection{Results}\label{sec:results}

\subsubsection{Level 2 Results.}

Table~\ref{tab:level2} summarizes the Level~2 results. The \qty{500}{\watt}
configuration, which achieved the best energy efficiency, was used for the
Green500 submission, while the \qty{700}{\watt} configuration, with the highest
absolute performance, was used for the Top500 submission.

\begin{figure}[ht]
  \centering
  \includegraphics{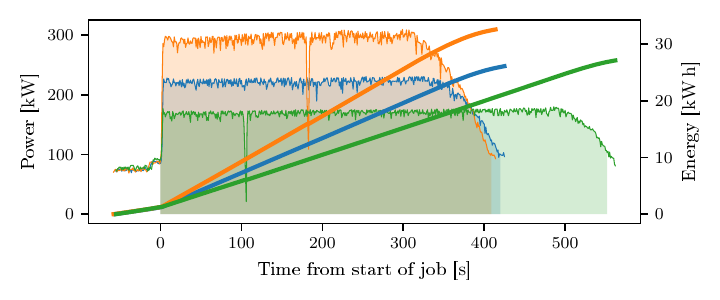}
  \caption{Power measured (thin lines) from PDUs and corresponding energies
    (thick lines) for 3 values of GPU power capping, \qty{350}{\watt} (green),
    \qty{500}{\watt} (blue), \qty{700}{\watt} (orange). Shaded regions
    correspond to the core part of HPL.}\label{fig:green500-lvl2}
\end{figure}

Since the \emph{Rmax} values were not as high as expected, a campaign of node
characterization was launched. The main results obtained from this campaign show
to what extent single nodes sustain HPL runs.

\subsubsection{Node Variability and Thermal Throttling}

\begin{figure}[ht]
  \centering
  \includegraphics{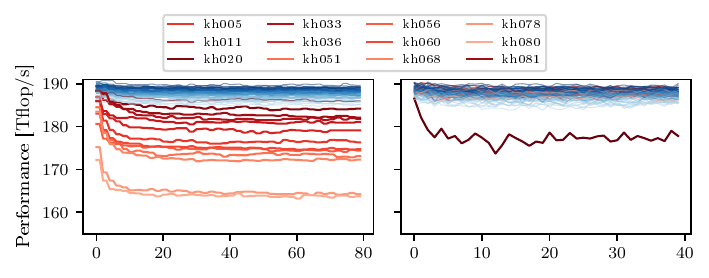}
  \caption{Consecutive runs of HPL, before (left) and after (right) vendor
    support intervention on the GPUs. The red thicker line corresponds to thermal
    throttling, the blue lines correspond to the nodes that behave properly}\label{fig:single-nodes}
\end{figure}

Prior to the Level~3 campaign, HPL was run 40 times on each node, once
allowing the GPU to cooldown between runs and once without. Several nodes
exhibited significant performance degradation over successive runs without
cooldown (\figref{fig:single-nodes}, left). While healthy nodes sustained
approximately \qty{189}{\tera\flop\per\second} consistently, the affected nodes
showed a significant performance drop after 2 to 3 runs. This is indicative of
thermal throttling under sustained GPU load. The fact that not all nodes exhibit
the same behavior was a sign of a potential hardware issue.

Investigation with the hardware vendor identified defective thermal pads on the
H100 GPUs of the affected nodes, which prevented adequate heat dissipation.
After replacement of the thermal pads by the vendor, the single-node tests were
repeated (\figref{fig:single-nodes}, right). The post-correction results
confirmed that the affected nodes recovered to nominal performance levels,
except for one node that developed issues only during the second campaign.

\subsubsection{Level 3 Results.}

\begin{table}[ht]
  \centering
  \caption{Level~3 results (August 2025, 80 nodes). PM = power meters.}
  \label{tab:level3}
  \begin{tabular}{lrrrr}
    \hline
    Power cap & Perf. & Core Energy & Core Avg. Power & Energy Efficiency \\
              & (\unit{\peta\flop\per\second}) & (\unit{\kilo\watth}) & (\unit{\kilo\watt}) & (\unit{{\giga\flop\per\second}\per\watt}) \\
    \hline
    \qty{700}{\watt} & 12.31 & 32.3 & 291.9 & 42.169 \\
    \qty{450}{\watt} & 10.99 & 25.6 & 207.4 & 52.979 \\
    \hline
  \end{tabular}
\end{table}

\begin{figure}[ht]
  \centering
  \includegraphics{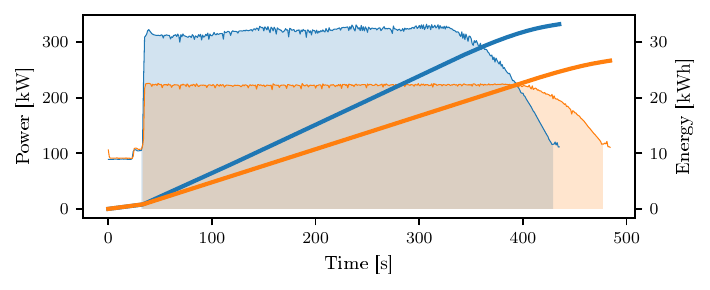}
  \caption{Energy measured from power meters and corresponding power for 2
    values of GPU power capping, \qty{450}{\watt} (orange) and \qty{700}{\watt}
    (blue). Shaded regions correspond to the core part of HPL.
  }\label{fig:green500-lvl3}
\end{figure}

Table~\ref{tab:level3} presents the Level~3 results submitted for the Top500 and
Green500 lists. The average power values are higher than those from the Level~2
PDUs for three reasons: more subsystems are connected to the power rails
(cooling fans and L40s extension network); integration errors differ between PDU
sampling at \qty{1}{\hertz} and power-meter energy integrated from data measured
at above \qty{5}{\kilo\hertz}; and PDU noise from Grafana
aggregation, visible in \figref{fig:green500-lvl2} and confirmed by comparison
with the power meter data in \figref{fig:green500-lvl3}.

It should be noted that the power meters had no network access to an NTP server
and required manual clock synchronization via a laptop. Despite this, clock
offsets between meters were corrected in post-processing by correlating the
power line frequency, which is identical across all rails. Residual offsets were
small: both PQ-50 meters were in sync, the PQ-150 was \qty{2}{\second} behind
and the Fluke~1738 was \qty{1.88}{\second} behind.

\subsubsection{Energy Efficiency Sweet Spot}
\begin{figure}[ht]
  \centering
  \includegraphics{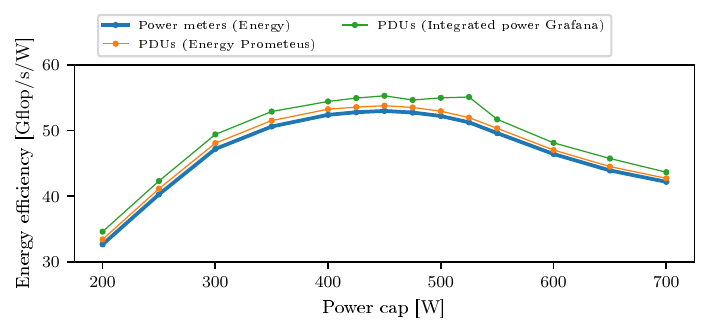}
  \caption{Energy efficiency for different GPU power
    capping.}\label{fig:energy-efficiency}
\end{figure}

The Level~2 campaign tested only three power capping values. During the
Level~3 campaign, this was extended to all \qty{50}{\watt} increments from
\qty{200}{\watt} to \qty{700}{\watt}, with additional \qty{25}{\watt} steps
between \qty{400}{\watt} and \qty{575}{\watt} to refine the sweep.
\figref{fig:energy-efficiency} shows the resulting energy efficiency curve,
together with Level~2-equivalent PDU values and a direct Prometheus measurement
for comparison.

The optimal energy efficiency is obtained for a capping of \qty{450}{\watt} with
a loss of $11\%$ of performance but a reduction of $20\%$ in energy consumption;
this configuration was selected for the Level~3 Green500 submission.
These results confirm what was already observed for example in
\cite{karimi_exploring_2025}, that energy savings can be achieved at low
performance losses through GPU power capping.

\section{Moving a step forward: Carbon footprint}\label{carbon}
The energy efficiency metric requested by the Green500 ranking offers a
complementary ranking to the performance-based TOP500 ranking. However, looking
solely at energy optimization obscures the overall environmental impact
generated by HPC clusters.

This section estimates the carbon footprint of \kuma, acknowledging that a full
assessment would require a multi-criteria life-cycle analysis beyond
$\textrm{CO}_2$ alone.

This study follows MITSI (Methodology for IT Services environmental Impact
assessment) \cite{mitsi-methodology}. This methodology follows the Life Cycle
Analysis framework, estimating the carbon footprint on four stages:
manufacturing (including abiotic resources extraction), distribution, usage and
end-of-life. We encompass the manufacturing, distribution and end-of-life phases
in what is commonly named embodied emissions.

\subsection{Usage emissions}

The electrical carbon footprint of data centers varies according to the time of
day, season, and geographic scope. For this study, we use a yearly national
average, which smooths out these variations. Such data are available from
platforms such as Electricity Maps\cite{electricitymaps}; for Switzerland, we
use the values provided by the Swiss association of electricity providers
(AES)\cite{aes25}: \qty{90}{{\gCOeq}\per\kilo\watth} for 2024 and
\qty{57}{{\gCOeq}\per\kilo\watth} for 2025.

As recommended by MITSI, the datacenter infrastructure electrical consumption
(cooling, lights, UPS overheads, etc.) is added to the usage emissions,
estimated using the Power Usage Effectiveness (PUE). The EPFL datacenter PUE is
estimated at 1.3, meaning \qty{0.3}{\kilo\watth} of datacenter infrastructure
overheads per \qty{1}{\kilo\watth} of IT load.

\subsection{Embodied emissions}

The embodied emissions cover the \kuma configuration used for the Green500
submission (Section~\ref{sec:setup}): the 80 H100 nodes, administration servers,
InfiniBand network, storage, and spare parts.

To get emission factors for each element composing the \kuma cluster, the
Resilio database \cite{resiliodb} was used. This database offers one of the most
up-to-date environmental database for ICT equipment and services, and has the
advantage to be freely available for academic users. Following MITSI with
Resilio data, we estimate a total footprint for the embodied emissions at
\qty{387.2}{\tCOeq}.

To add the embodied emissions to usage emissions, this total footprint was
allocated based on the total lifetime of the cluster, which has been committed
to 6 years with a 100\% usage. This gives us an embodied cost per second of
\qty{2.1}{{\gCOeq}\per\second}.

For \kuma, the embodied emission per GPU-hour is
\qty{23.0}{\gCOeq\per\gpuhour}. This is significant: the commonly used
GPU-hour metric encompasses only the usage phase, which, depending on the
electrical carbon footprint, can vary between approximately
\qty{40}{\gCOeq\per\gpuhour} and \qty{500}{\gCOeq\per\gpuhour}, meaning
embodied emissions represent between 10\% and 60\% of usage emissions.

The embodied emissions of the datacenter building and cooling infrastructure
were not included, as these are a minor contribution to total datacenter
footprint\cite{alissa25,bux25}.

\subsection{Results}

\begin{figure}[htb]
  \centering
  \includegraphics{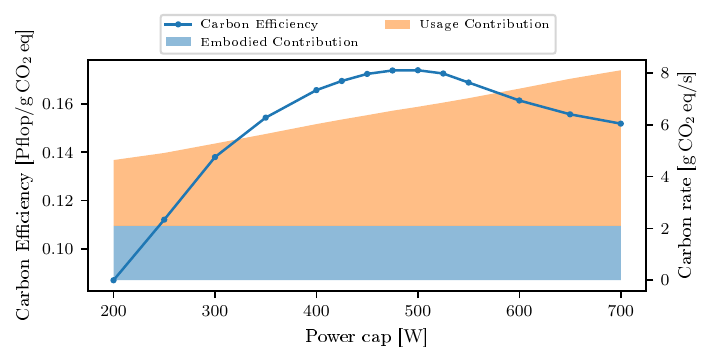}
  \caption{Carbon efficiency for different GPU power
    capping for the Level~3 measures.}\label{fig:carbon-efficiency}
\end{figure}

The methodology was applied to the Level~3 measurements to assess the carbon
emissions of the \kuma cluster, both for embodied and usage emissions. The
results are shown in \figref{fig:carbon-efficiency}.

As expected, the embodied contribution is constant, as it is modeled as a fixed
carbon rate independent of the power cap. On the other hand, the usage
contribution scales roughly linearly with the power cap, as higher GPU power
directly increases the average power draw. This contribution is therefore low at
\qty{200}{\watt} and high at \qty{700}{\watt}.

\begin{equation}\label{eq:carbon-efficiency}
  C_\mathrm{eff} = \frac{Perf}{C_{\mathrm{embodied}} + C_{\mathrm{usage}}} = \frac{Perf}{C_{\mathrm{embodied}} + (\bar{P} \cdot CI \cdot PUE)}
\end{equation}

We now analyze the carbon efficiency $C_\mathrm{eff}$ as defined in
Eq.\ref{eq:carbon-efficiency}, where $Perf$ is the performance in
\unit{\flop\per\second}, $C_{\mathrm{embodied}}$ and $C_{\mathrm{usage}}$ the
embodied and usage carbon rates, in \unit{\gCOeq\per\second}, $\bar{P}$ the
average power per run in \unit{\watt}, $CI$ the carbon intensity in
\unit{\gCOeq\per\kilo\watth}. Taking as electricity carbon footprint
\qty{57}{{\gCOeq}\per\kilo\watth}---2025 Switzerland value---and as embodied
emissions \qty{2.1}{{\gCOeq}\per\second}---since our cluster lifetime is
committed to 6 years---a similar behavior is observed between the carbon
efficiency and the energy efficiency when power capping, but with a maximum of
carbon efficiency observed at a power capping of \qty{500}{\watt} instead of
\qty{450}{\watt} for energy efficiency. Indeed, in such condition (low carbon
electrical footprint and 6 years cluster lifetime), the embodied emissions have
an important contribution: to reduce this contribution, the cluster lifetime should be
extended as much as possible, whereas to reduce usage contribution and maximizing
energy efficiency, it should run as efficiently as possible.

\begin{figure}[htb]
  \centering
  \includegraphics{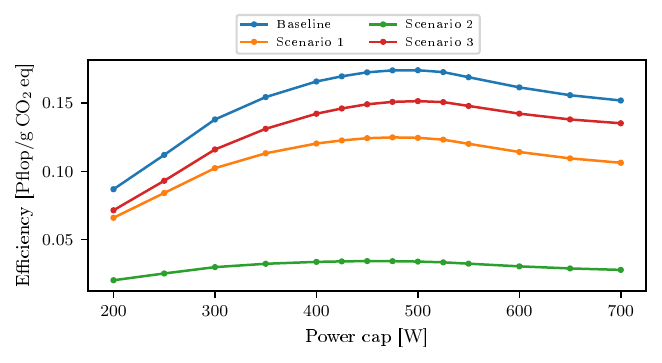}
  \caption{Carbon efficiency for the 4 scenarios (Section \ref{4-scenarios}).}\label{fig:carbon-efficiency-sensitivity}
\end{figure}

We now consider three alternative scenarios varying the energy mix and cluster
lifetime. Scenario~1 reflects our 2024 measurement conditions
(\qty{90}{{\gCOeq}\per\kilo\watth}, 6~years). Scenario~2 places the cluster in a
region with a higher-carbon electrical mix (\qty{400}{{\gCOeq}\per\kilo\watth},
6~years). Scenario~3 reduces the cluster lifetime to 4~years, raising the
embodied cost per second to \qty{3.1}{{\gCOeq}\per\second}
(\qty{57}{{\gCOeq}\per\kilo\watth}, 4~years)\label{4-scenarios}.

The four scenarios (baseline, 1, 2, 3) are shown in
\figref{fig:carbon-efficiency-sensitivity}. Whether we consider Swiss electrical
mix on 2024 or 2025, or we reduce the cluster lifetime to 4 years (adding then a
higher importance to embodied emissions), the power capping that maximizes the
carbon efficiency stays at \qty{500}{\watt}. It is only when considering a high
energy mix around \qty{400}{\gCOeq\per\kilo\watth} that usage contribution has
much more impact than the cluster lifetime. There, the carbon efficiency is much
closer to the energy efficiency.

\section{Discussions}\label{discussion}

\subsection{On the importance of taking into account average vs.\ measured power
  estimation}

MITSI estimates usage emissions based on the average power
consumption of the cluster over a full year of operation. This yearly average
smooths out the variability between workloads and idle periods, and is
well-suited for an annual carbon footprint assessment of the data center as a
whole. However, it does not reflect the actual energy consumed during a specific
computation.

In this paper, the carbon efficiency presented in Section~\ref{carbon} was
computed from the actual power measured during the HPL runs, rather than from
the yearly average. This distinction matters: the HPL benchmark stresses the
cluster close to its peak power draw, making it unrepresentative of typical
usage. As a consequence, the carbon efficiency figures reported here should be
understood as the efficiency of the cluster \emph{during the benchmark}, not
during normal operation.

\subsection{Toward a Carbon500 ranking}

As discussed in previous sections, MITSI assesses the carbon footprint of IT
services, during their whole lifespan, from manufacturing and distribution to
usage and end-of-life. Applied to HPC clusters from Top500, a new ranking could
be derived, named for instance "Carbon500", but it raises several challenges. A
major contribution in such a "Carbon500" ranking would be the electrical carbon
footprint, which is largely based on the location of the considered HPC cluster.
Even if energy efficiency of data center does play a role in usage emissions,
the carbon efficiency is mainly driven by the electrical carbon footprint.
Different strategies are possible to assess the electrical carbon footprint,
which could blur the comparison between different HPC clusters. Moreover, the
annual electrical carbon footprint can be known only the following year, meaning
that the cluster carbon footprints could only be computed up to a year after the
measurements are achieved.

Regarding the measurement of the embodied part, the data source is still too
fragmented. Many parameters are only declarative and cannot be checked. This is
particularly true for the data center's PUE, and the server type (number of GPUs,
network parameters, etc.) for instance. This would complicate the validation of
this ranking.

Nevertheless, focusing the ranking on carbon impact enables HPC clusters to be
compared on their environmental impact, even if only from a carbon perspective.
It has two positive impacts: it helps estimate the environmental impact of
computation, especially from the scientific community, and it enables choosing
the most environmentally friendly clusters on which large scientific projects
are run. It then helps reduce their environmental impact
and promote low-carbon research.

\section{Conclusion}

This study reports two Green500 submissions (Level~2 in 2024, Level~3 in 2025)
and assesses the full carbon footprint of the \kuma cluster, covering both
embodied and usage emissions. The Green500 energy efficiency optimum is found at
a GPU power cap of \qty{450}{\watt}. Results differ slightly depending on
whether we consider the energy efficiency or the carbon emission efficiency,
where for the latter an optimum is found closer to \qty{500}{\watt} for HPC
clusters running in a low-carbon electrical grid such as Switzerland. Other
benchmarks, closer to real AI loads, should also be tested as they become a
common run in HPC infrastructures.

Green500 is already an improvement with respect to Top500, assessing the
HPC cluster in regards of energy efficiency. But to encompass environmental
impacts, this is not enough. Another ranking, focused on carbon emissions, that
we propose to name Carbon500, will also integrate the fluctuations of electrical
mix, highly dependent on HPC locations and years. This would be a way to enhance
motivation to highlight low-carbon-emission data centers---where scientific
computation could then be located---and strengthen ambitious policies to reduce
the environmental footprint of electricity production.

Finally, carbon assessment alone is not enough to capture the full environmental
impact of HPC; a multi-criteria life-cycle analysis encompassing abiotic
resource depletion, water and land use, freshwater eco-toxicity, and fossil
resource usage is needed to avoid erroneous conclusions\cite{resilio_swiss}.

\begin{credits}
  \subsubsection{\ackname} MITSI was made thanks to the contribution of the
  Swiss Centre of Excellence on Net-Zero Emissions (SCENE \cite{scene}).

  Anthropic Claude was used as an AI assistant to help write this article.

\subsubsection{\discintname}
The authors have no competing interests to declare that are
relevant to the content of this article.
\end{credits}

%
%
\bibliographystyle{splncs04}
\bibliography{biblio}

@article{feng-2007,
  author={Feng, Wu-chun and Cameron, Kirk},
  journal={Computer}, 
  title={The Green500 List: Encouraging Sustainable Supercomputing}, 
  year={2007},
  volume={40},
  number={12},
  pages={50-55},
  doi={10.1109/MC.2007.445}}

@misc{top500-1993,
  author = {{TOP500.org}},
  title = {TOP500},
  year = 1993,
  url = {https://top500.org/lists/top500/},
}

@misc{green500-2013,
  author = {{TOP500.org}},
  title = {Green500},
  year = 2013,
  url = {https://top500.org/lists/green500/},
}

@misc{HPL-mixed,
  author = {},
  title = {HPL-MXP mixed-precision benchmark},
  url = {https://hpl-mxp.org/}
}

@misc{SCITAS-Kuma,
  author = {{EPFL SCITAS}},
  title = {EPFL SCITAS Kuma},
  url = {https://www.epfl.ch/research/facilities/scitas/hardware/kuma/}
}

@article{mccalpin1995stream,
  author = {John D. McCalpin},
  title = {Memory Bandwidth and Machine Balance in Current High Performance
                  Computers},
  journal = {IEEE Computer Society Technical Committee on Computer Architecture
                  (TCCA) Newsletter},
  year = {1995},
  pages = {19--25},
  month = dec
}

@techreport{heroux2013hpcg,
  title={HPCG benchmark technical specification},
  author={Heroux, Michael Allen and Dongarra, Jack J. and Luszczek, Piotr},
  year={2013},
  institution={Sandia National Lab.(SNL-NM), Albuquerque, NM (United States)}
}

@inproceedings{kudo2020implementation,
  title={Implementation and numerical techniques for one EFlop/s HPL-AI benchmark on Fugaku},
  author={Kudo, Shuhei and Nitadori, Keigo and Ina, Takuya and Imamura, Toshiyuki},
  booktitle={2020 IEEE/ACM 11th Workshop on Latest Advances in Scalable Algorithms for Large-Scale Systems (ScalA)},
  pages={69--76},
  year={2020},
  organization={IEEE},
  doi={10.1109/ScalA51936.2020.00014}
}

@book{dongarra1979linpack,
  title={LINPACK users' guide},
  author={Dongarra, Jack J. and Moler, Cleve Barry and Bunch, James R and Stewart, Gilbert W},
  year={1979},
  publisher={SIAM}
}

@inproceedings{li2023toward,
  title={Toward sustainable hpc: Carbon footprint estimation and environmental implications of hpc systems},
  author={Li, Baolin and Basu Roy, Rohan and Wang, Daniel and Samsi, Siddharth and Gadepally, Vijay and Tiwari, Devesh},
  booktitle={Proceedings of the international conference for high performance computing, networking, storage and analysis},
  pages={1--15},
  year={2023},
  doi={10.1145/3581784.3607035}
}

@article{dongarra2025hpl,
  title={HPL-MxP benchmark: Mixed-precision algorithms, iterative refinement, and scalable data generation},
  author={Dongarra, Jack J. and Luszczek, Piotr},
  journal={The International Journal of High Performance Computing Applications},
  pages={10943420251382476},
  year={2025},
  publisher={SAGE Publications Sage UK: London, England},
  doi={10.1177/10943420251382476}
}

@article{strohmaier2015top500,
  title={The TOP500 list and progress in high-performance computing},
  author={Strohmaier, Erich and Meuer, Hans W and Dongarra, Jack J. and Simon, Horst D},
  journal={Computer},
  volume=48,
  number=11,
  pages={42--49},
  year=2015,
  doi={10.1109/MC.2015.338}
}

@inproceedings{kline2016holistically,
  title={Holistically evaluating the environmental impacts in modern computing systems},
  author={Kline, Donald and Parshook, Nikolas and Ge, Xiaoyu and Brunvand, Erik and Melhem, Rami and Chrysanthis, Panos K and Jones, Alex K},
  booktitle={2016 Seventh International Green and Sustainable Computing Conference (IGSC)},
  pages={1--8},
  year={2016},
  organization={IEEE},
  doi={10.1109/IGCC.2016.7892605}
}

@inproceedings{gupta2021chasing,
  title={Chasing carbon: The elusive environmental footprint of computing},
  author={Gupta, Udit and Kim, Young Geun and Lee, Sylvia and Tse, Jordan and Lee, Hsien-Hsin S and Wei, Gu-Yeon and Brooks, David and Wu, Carole-Jean},
  booktitle={2021 IEEE International Symposium on High-Performance Computer Architecture (HPCA)},
  pages={854--867},
  year={2021},
  organization={IEEE},
  doi={10.1109/HPCA51647.2021.00076}
}

@article{adhinarayanan2025looking,
  title={Looking Back to Look Forward: 15 Years of the Green500},
  author={Adhinarayanan, Vignesh and Feng, Wu-chun},
  journal={Computer},
  volume={58},
  number={1},
  pages={76--86},
  year={2025},
  publisher={IEEE},
  doi={10.1109/MC.2023.3333316}
}

@techreport{green500-methodology,
  title={Energy Efficient High Performance Computing Power Measurement Methodology},
  author={{The Green500} and {Top500} and {The Green Grid} and {EEHPC WG}},
  year={2014},
  note={Version 2.0 RC 1.0},
  url={https://www.top500.org/static/media/uploads/methodology-2.0rc1.pdf}
}

@techreport{mitsi-methodology,
  title={{MITSI} - Methodology for IT Services environmental Impact assessment},
  author={Paolini, Julia and Cubero-Castan, Manuel and Ouvrard, Xavier and Peon Quirós, Miguel and Richart, Nicolas},
  year={2026},
  url={https://infoscience.epfl.ch/handle/20.500.14299/261022},
  note={Version 1.0 - to be published soon}
}

@techreport{resiliodb,
  title={Environmental impacts of IT equipment - A comparative analysis - Resilio Database and ‘Product Carbon Footprint’ sheets},
  author={Aubet, Louise and Violet, Benjamin},
  year={2024},
  url={https://resilio-solutions.com/files/news/publications/comparative-analysis/Resilio_Aguaro_Whitepaper_RDB_vs_PCF_20241025_web_en.pdf}
}

@article{alissa25,
  title={Using life cycle assessment to drive innovation for sustainable cool clouds},
  author={Alissa, Husam and Nick, Teresa and Raniwala, Ashish and et al.},
  journal={Nature},
  volume={641},
  number={8062},
  pages={331--338},
  year={2025},
  doi={10.1038/s41586-025-08832-3}
}

@article{bux25,
  title={A critical analysis of global warming potential of data centers in the digital era},
  author={Bux, Christian and Rana, Roberto Leonardo and Lombardi, Mariarosaria and Giungato, Pasquale and Tricase, Caterina},
  journal={The International Journal of Life Cycle Assessment},
  volume={30},
  number={11},
  pages={2390--2402},
  year={2025},
  doi={10.1007/s11367-024-02419-2}
}

@misc{electricitymaps,
  author = {{Electricity Maps Aps}},
  title = {Electricity Maps Portal},
  url = {https://app.electricitymaps.com/},
}

@techreport{aes25,
  title={Teneur en $\mathrm{CO}_2$ du mix électrique suisse d’ici 2050},
  author={{Association des entreprises électriques suisses AES}},
  year={2026},
  url={https://www.strom.ch/fr/document/teneur-en-co2-du-mix-electrique-suisse-dici-2050-2026-0}
}

@techreport{resilio_swiss,
  title={Environmental impacts of the ICT sector in Switzerland},
  author={Resilio},
  year={2025},
  note={Version 1.0 - September 2025}
}

@inproceedings{karimi_exploring_2025,
  location = {Atlanta, {GA}, {USA}},
  title = {Exploring the Frontiers of Energy Efficiency using Power Management at System Scale},
  booktitle= {{Proceedings of the SC '24 Workshops of the International Conference on High Performance Computing, Network, Storage, and Analysis}},
  isbn = {979-8-3503-5554-3},
  doi = {10.1109/SCW63240.2024.00230},
  series = {{SC}-W '24},
  pages = {1835--1844},
  publisher = {{IEEE} Press},
  author = {Karimi, Ahmad Maroof and Maiterth, Matthias and Shin, Woong and Sattar, Naw Safrin and Lu, Hao and Wang, Feiyi},
  date = {2025-02-11},
  year = {2025},
}

@misc{scene,
  title={SCENE Project},
  author={{SCENE Project}},
  url={https://www.scene-project.ch/},
}

\end{document}